\documentclass[12pt,preprint]{aastex}
\usepackage{natbib}
\usepackage{ifthen}
\newcounter{address}
\newcommand{\latin}[1]{\textit{#1}}
\newcommand{\ie}{\latin{i.e.}}
\newcommand{\eg}{\latin{e.g.}}
\newcommand{\cf}{\latin{cf.}}

\newcommand{\rp}{r_\mathrm{p}}
\newcommand{\rpmin}{r_\mathrm{p,min}}
\newcommand{\Dcl}{D_\mathrm{cl}}
\newcommand{\Rvir}{R_\mathrm{vir}}
\newcommand{\Halpha}{\ensuremath{\mathrm{H}\alpha}}
\newcommand{\Hdelta}{\ensuremath{\mathrm{H}\delta}}
\newcommand{\OII}{\ensuremath{[\mathrm{O}\,II]}}

\begin{document}
\title{
  What triggers galaxy transformations?\\
  The environments of post-starburst galaxies
}
\author{
  David~W.~Hogg\altaffilmark{\ref{NYU},\ref{email}},
  Morad~Masjedi\altaffilmark{\ref{NYU}},
  Andreas~A.~Berlind\altaffilmark{\ref{NYU}},
  Michael~R.~Blanton\altaffilmark{\ref{NYU}},
  Alejandro~D.~Quintero\altaffilmark{\ref{NYU}},
  J. Brinkmann\altaffilmark{\ref{APO}}
}

\setcounter{address}{1}
\altaffiltext{\theaddress}{\stepcounter{address}\label{NYU}
Center for Cosmology and Particle Physics, Department of Physics, New
York University, 4 Washington Place, New York, NY 10003}
\altaffiltext{\theaddress}{\stepcounter{address}\label{email}
To whom correspondence should be addressed: \texttt{david.hogg@nyu.edu}}
\altaffiltext{\theaddress}{\stepcounter{address}\label{APO}
Apache Point Observatory, 2001 Apache Point Road,
P.O. Box 59, Sunspot, NM 88349}

\begin{abstract}
Star-formation history is strongly related to environment; the most
massive and least star-forming galaxies reside in the highest density
environments.  There are now good observational reasons to believe
that the progenitors of these red galaxies have undergone starbursts,
followed by a post-starburst phase.  Post-starburst (``K+A'' or
``E+A'') galaxies appear in the SDSS visible spectroscopic data by
showing an excess of A star light (relative to K giant light) but
deficient $\Halpha$ line emission.  We investigate the environments of
these galaxies by measuring \textsl{(1)}~number densities in
$8\,h^{-1}~\mathrm{Mpc}$ radius comoving spheres,
\textsl{(2)}~transverse distances to nearest Virgo-like galaxy
clusters, and \textsl{(3)}~transverse distances to nearest
luminous-galaxy neighbors.  We compare the post-starburst galaxies to
currently star-forming galaxies identified solely by A-star excess or
$\Halpha$ emission.  We find that post-starburst galaxies are in the
same kinds of environments as star-forming galaxies; this is our
``null hypothesis''.  More importantly, we find that at each value of
the A-star excess, the star-forming and post-starburst galaxies lie in
very similar distributions of environment.  Other studies finding
similar results have argued that galaxy transformations occur slowly
(time scales $>1$~Gyr), but this is at odds with the observational
evidence that red galaxies are formed via starbursts.  The only
deviations from our null hypothesis are barely significant: a slight
deficit of post-starburst galaxies (relative to the star-forming
population) in very low-density regions, a small excess inside the
virial radii of clusters, and a slight excess with nearby neighbors.
None of these effects is strong enough to make the post-starburst
galaxies a high-density phenomenon, or to argue that the starburst
events are primarily triggered by external tidal impulses (\eg, from
close passages of massive galaxies).  The small excess inside cluster
virial radii suggests that some post-starbursts are triggered by
interactions with the intracluster medium, but this represents a very
small fraction of all post-starburst galaxies.
\end{abstract}


\section{Introduction}

How do old, dead, early-type galaxies form?  There are two strong
arguments that bulge-dominated galaxies (\eg, ellipticals,
lenticulars, and very early-type spirals) have progenitors that went
through a starburst phase.  The first is that bulge-dominated galaxies
show enhancements in $\alpha$-type elements over the Solar chemical
abundance mix \citep[\eg, ][]{worthey98a, eisenstein03b}.  These
abundance patterns are naturally produced when star formation---or the
last phase of star formation---has occured in a burst too rapid to
allow recycling of the elements ejected by Type~1a supernovae.  The
second argument is that the post-starburst galaxies identified
spectrally in large surveys of the Local Universe have the colors,
surface brightnesses, and radial profiles that would be expected if
they are to evolve passively into new bulge-dominated galaxies
\citep{quintero04a}; they cannot evolve passively into disk-dominated
or other late types.  Starburst origin is also supported indirectly by
the uniformity seen in early-type galaxies' stellar populations
\citep[\eg, ][]{eisenstein03b} and their lack of large resevoirs of
cold gas and dust.

Post-starburst galaxies are identified spectroscopically by having a
large contribution to their spectral energy distribution from A
stars---\ie, new stars must have formed within the last $\sim
1~\mathrm{Gyr}$---but no, or very little, contribution from O and B
stars---\ie, no new stars have formed within the last $\sim
0.01~\mathrm{Gyr}$.  In practice, these ``K+A'' or ``E+A''
galaxies\footnote{The terminology ``K+A'' is to be preferred to
``E+A'' because the identification is spectral, not morphological, and
``K'' and ``A'' name spectral types.  ``E'' names a morphological type
\citep{franx93a, dressler99a}.} \citep{dressler83a, zabludoff96a,
dressler99a} are identified by having strong Balmer absorption or blue
continua (A-star indicators) but no $\Halpha$ or [O\,II] emission
lines (O and B-star indicators).  Importantly, because A stars have
known lifetimes, the evolution of the population can be ``timed'' and
rates computed; we find that of order 1~percent of the galaxy
population is going through this phase each Gyr at $z\sim 0.1$
\citep{quintero04a}.

What is not known is what precedes or triggers the (necessarily rapid)
truncation of star-formation.  Is it an external event, such as a
tidal impulse, accretion event, or major merger?  Or is it a purely
internal event, such as an AGN flare or the abrupt exhaustion of
star-formation fuel?  Either way, since disk-dominated galaxies are
the galaxies that contain young stars and the cold-gas fuel to make
more, post-starburst galaxies must lie on some kind of evolutionary
pathway between the disk-dominated and bulge-dominated populations.

The distribution of galaxy star-formation rates is a strong function
of environment, with much lower star-formation rates in higher density
regions \citep[\eg, ][]{kennicutt83a, hashimoto98a, balogh01a,
martinez02a, lewis02a, gomez03a, blanton03d, hogg03b, hogg04a,
kauffmann04a, blanton05b, quintero06a}.  Although it is customary to
think of this as being a consequence of the morphology--density
relation \citep{dressler80a, postman84a}, in fact recent studies with
large samples have shown that in fact the star-formation--environment
relation has more explanatory or informative power than the
morphology--density relation, at least with the morphological proxies
currently available for large samples \citep{hashimoto98a,
hashimoto99a, kauffmann04a, blanton05b, quintero06a}.  How is the
information about environment transmitted to galaxy star-formation
activity?  Are there violent events when galaxies fall into high
density regions?  Or do the galaxies reduce their star-formation rates
gradually as they find themselves in denser and denser environments?
Studies of the (strong, observed) evolution of the fraction of
galaxies in clusters that are blue (often called the ``Butcher--Oemler
effect''; \citealt{butcher78a}) have generally concluded that this
evolution is gradual, at least when compared to the lifetimes of A
stars \citep{poggianti99a, balogh00a, kodama01a}, although there are
certainly some galaxies in clusters that appear to be undergoing rapid
evolution \citep[\eg, ][]{vogt04a, yang04a}.  It is not clear how to
reconcile this conclusion of gradual evolution with the conclusion
(mentioned above) that a typical early-type galaxy has undergone a
massive starburst in its past.

Galaxy star-formation rates are also evolving very rapidly with
redshift in the field; this result comes from many different
techniques at many different wavelengths \citep[\eg, ][]{lilly96,
hammer97, rowan-robinson97, hogg98o2lf, tresse98, cowie99, flores99,
mobasher99, haarsma00, juneau05a, schiminovich05a}.  This is usually
imagined as being related not to infall into dense regions but rather
to the supply of cold gas.  On the other hand, since gravitational
clustering brings galaxies into more and more dense environments with
cosmic time, this might not be unrelated to the
star-formation--environment relation and the Butcher--Oemler effect.

At the same time as star formation in the Universe is declining, the
total density of stellar mass on the ``red sequence'' of early-type or
bulge-dominated galaxies is increasing \citep{bell04a, blanton06a,
faber05a, brown06a}.  If blue galaxies transform into red galaxies via
a post-starburst phase, as we believe they must, then this evolution
of the red sequence must be quantitatively matched with an evolving
population of post-starburst galaxies.

As transition objects between the star-forming, disk-dominated and
dead, bulge-dominated populations, the post-starburst galaxies could
in principle have the environmental characteristics of either.
Originally, K+A galaxies were found in high-density regions
\citep{dressler83a, couch87a}, and thought to be a ``cluster''
population.  Of course the early searches for such galaxies were made
in cluster fields.  Once systematic searches for K+A galaxies were
made in large redshift surveys, it was found that they are not
particularly concentrated in clusters or high density regions, but
rather live in a wide range of environments \citep[][Helmboldt et~al.,
in preparation]{zabludoff96a, quintero04a, blake04a, goto05a}.  In the
large SDSS and 2dFGRS samples, it can be shown that the mean
environment \citep[][Helmboldt et al., in preparation]{quintero04a}
and distribution of environments \citep{blake04a} are both similar to
those of spiral or disk-dominated galaxies.

The environments of disk-dominated galaxies---isolation and small
groups, where the virialized mass has a similar velocity dispersion to
the contained galaxies---are the best environments in the Local
Universe to find galaxy--galaxy mergers, which are the top candidates
for triggers for the post-starburst galaxies.  After all, the major
mergers observed in the Local Universe are all associated with very
high star-formation rates, and major mergers are expected to disrupt
disks and leave behind the dynamically hot stellar orbits
characteristic of the bulge-dominated population.

With a sample of more than $10^3$ K+A galaxies \citep{quintero04a}, we
are in a position to ask much more detailed questions about the range
of environments in which they lie, and the relationships between
environmental and star-formation properties.  That is the purpose of
this \textsl{Article}, with the goal of constraining the possible
triggering mechanisms for this very important galaxy population.

In what follows, a cosmological world model with
$(\Omega_\mathrm{M},\Omega_\mathrm{\Lambda})=(0.3,0.7)$ is adopted,
and the Hubble constant is parameterized
$H_0=100\,h~\mathrm{km\,s^{-1}\,Mpc^{-1}}$, for the purposes of
calculating distances\citep[\eg,][]{hogg99cosm}.

\section{Data}

The SDSS is taking $ugriz$ CCD imaging of $10^4~\mathrm{deg^2}$ of the
Northern Galactic sky, and, from that imaging, selecting $10^6$
targets for spectroscopy, most of them galaxies with
$r<17.77~\mathrm{mag}$ \citep{gunn98a, york00a, stoughton02a,
abazajian03a, abazajian04a}.

All the data processing, including astrometry \citep{pier03a}, source
identification, deblending and photometry \citep{lupton01a},
calibration \citep{fukugita96a, smith02a, ivezic04a}, spectroscopic target
selection \citep{eisenstein01a, strauss02a, richards02a}, spectroscopic
fiber placement \citep{blanton03a}, spectral data reduction and
analysis (Schlegel \& Burles, in preparation, Schlegel in preparation)
are performed with automated SDSS software.

Galaxy absolute magnitudes and colors are computed in fixed
bandpasses, using Galactic extinction corrections \citep{schlegel98a}
and $K$ corrections \citep[computed with \texttt{kcorrect
v1\_11};][]{blanton03b}.  They are $K$ corrected not to the redshift
$z=0$ observed bandpasses but to bluer bandpasses $^{0.1}g$, $^{0.1}r$
and $^{0.1}i$ ``made'' by shifting the SDSS $g$, $r$, and $i$
bandpasses to shorter wavelengths by a factor of 1.1
\citep[\cf,][]{blanton03b, blanton03d}.  This means that galaxies at
redshift $z=0.1$ (typical of the sample used here) have trivial (but
non-zero) $K$ corrections.

We capitalize on the extremely good spectrophotometric calibration of
the SDSS data and measure, for each galaxy, the excess light in each
fiber spectrum in the wavelength range $3800<\lambda<5400~\mbox{\AA}$
coming from A-type stars relative to K-type stars, normalized to the
mean spectrum of an old galaxy in the SDSS.  This measurement is
described elsewhere \citep{quintero04a}; briefly, we perform a linear
fit of the spectral section to a linear combination of the mean SDSS
old galaxy spectrum (the ``K'' spectrum) and the mean SDSS A-star
spectrum (the ``A'' spectrum) with the locations of possible emission
lines masked out.  The A-star excess is then the ratio $A/K$ of the
amplitudes of the two spectral components from the fit.  We also
measure the line flux and equivalent width (EW) of the $\Halpha$ line
using the ``K+A'' fit as a continuum model (which effectively removes
the absorption contribution to the flux at $\Halpha$).  These
measurements are performed exactly as described by us previously
\citep{quintero04a}.

\begin{figure}
\plotone{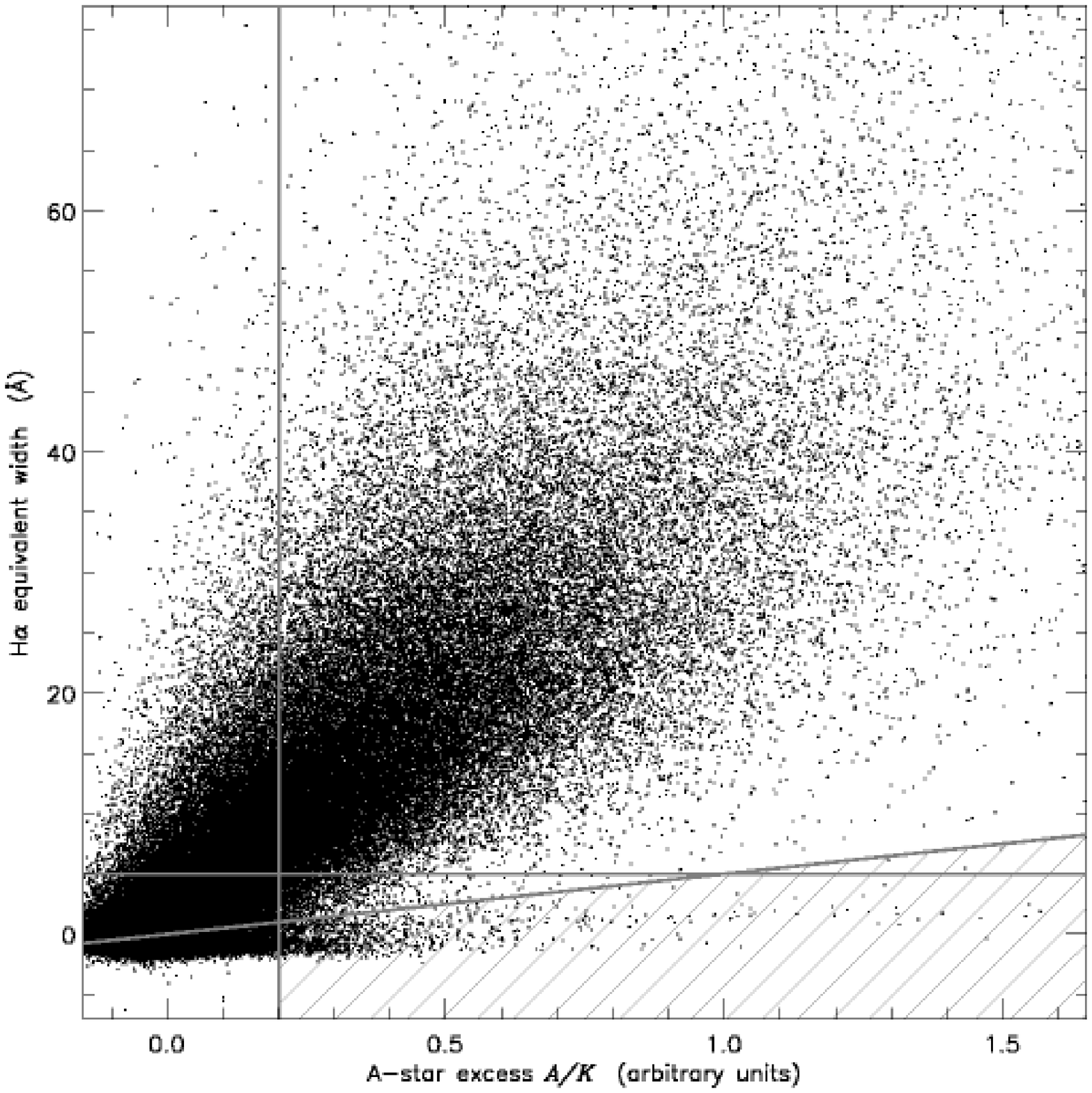}
\caption{The $\Halpha$~EW and A-star excess $[A/K]$ measurements for
  the sample galaxies.  Galaxies above the horizontal line are deemed
  ``star-forming'' by $\Halpha$~EW, galaxies right of the vertical line
  are deemed ``star-forming'' by A-star excess, and galaxies in the
  hatched region are deemed ``K+A''.  Galaxies below the sloped line
  are deemed ``low $\Halpha$''.\label{fig:quintero}}
\end{figure}

The ``units'' in which the A-star excess is measured are arbitrary,
but used here (where the units correspond to a luminosity ratio in the
above-mentioned wavelength band) they can be calibrated by their
relationship to $\Halpha$~EW; typically
\begin{equation}
\frac{A}{K} \sim \frac{\Halpha~\mathrm{EW}}{40~\mbox{\AA}}
\label{eqn:slope}
\end{equation}
\citep{quintero04a}.  For the purposes of what follows, we define
``low $\Halpha$'' galaxies to be those with $\Halpha$~EW less than
$1/8$ the value implied by equation (\ref{eqn:slope}).  We define
star-forming ``$\Halpha$ excess'' galaxies to be those with
$\Halpha~\mathrm{EW}>5.0~\mbox{\AA}$ and star-forming ``A-star
excess'' galaxies to be those with $(A/K)>0.2$ in our arbitrary units.
K+A galaxies, the subject of this study, are those that are ``low
$\Halpha$'' but ``A-star excess''; \ie, they are star-forming
according to the A-star excess, but not according to the $\Halpha$
emission.  These definitions are illustrated graphically in
Figure~\ref{fig:quintero}.

Often, K+A galaxies are found in surveys by their large $\Hdelta$
absorption EW, because that indicates A stars but little line emission
to dilute or fill the absorption trough \citep{zabludoff96a, blake04a,
goto05a}.  These studies usually use other emission lines (such as
\OII) to veto star-forming galaxies, but they are less general than
what is presented here, because they select only for post-starburst
galaxies dominated by the A stars with the strongest $\Hdelta$ lines,
and they do not separate post-starburst galaxies out of the main
galaxy population as cleanly as the method used here.  On the other
hand, the $\Hdelta$ method can be used with poorly calibrated spectra,
because it relies only on a local measure of the spectral shape around
the $\Hdelta$ line.  Fortunately, with the SDSS spectra,
spectrophotometric calibration is excellent \citep{quintero04a}.

In principle it is possible for an actively star-forming object to
appear post-starburst if the very young stars are fully
dust-enshrouded but migrate out of the dusty regions on timescales of
$< 1~\mathrm{Gyr}$ \citep{poggianti00a, quintero04a, blake04a}.
Though this is possible, even the dustiest starbursts known do emit
significant $\Halpha$, and indeed 20~cm radio observations of a sample
of 36 SDSS post-starburst candidates find no evidence for significant,
hidden star-formation \citep{goto04a}.

Around every galaxy a density $\rho_8$ is measured as described
elsewhere \citep{blanton03d, hogg03b}; briefly, it is a count of the
number of neighbors in the SDSS spectroscopic sample inside a
$8\,h^{-1}~\mathrm{Mpc}$ radius comoving sphere in comoving distance
space (with no correction for redshift distortions), divided by the
uniform-density predicted number, made from the galaxy luminosity
function \citep{blanton03c} and the SDSS window function, to make a
dimensionless density.  The sample used to infer $\rho_8$ is
flux-limited and not volume-limited, but the resulting overdensity
estimates have been shown to be redshift-independent in the median
\citep{blanton03d}.  The relatively large radius of
$8\,h^{-1}~\mathrm{Mpc}$ is chosen to provide good signal-to-noise per
object, although we have shown that there is in fact more information
on smaller scales \citep{blanton04c}.

For each galaxy we estimate a second environment measure by finding
the transverse distance $\Dcl$ to the nearest Virgo-like (or greater)
galaxy cluster that is within $1000~\mathrm{km\,s^{-1}}$ in radial
distance, divided by the virial radius $\Rvir$ of the cluster.  The
galaxy clusters are $\geq 10$-member clusters taken from a
friends-of-friends cluster catalog constructed from SDSS Main Sample
galaxies with absolute magnitudes $M_{^{0.1}r}<-19.9$~mag
\citep{berlind06a}.  The 10-member limit was chosen here because,
after converting $M_B$ approximately to $M_{^{0.1}r}$, we find about
13 galaxies in Virgo at this limit \citep{trentham02a}.  Cluster
abundance as a function of multiplicity is used to convert
multiplicity to mass \citep{berlind06a} and the mass is used to
compute a virial radius $\Rvir$ at which the cluster represents an
overdensity of 200.

On the smallest scales, we measure galaxy environment in a third way
by the transverse nearest-neighbor distance $\rpmin$.  This is defined
here to be the transverse proper distance $\rp$ to the closest
neighbor galaxy in the SDSS spectroscopic sample with $^{0.1}i$-band
absolute magnitude $M_{^{0.1}i}$ brighter than $-20.0~\mathrm{mag}$
and within $200~\mathrm{km\,s^{-1}}$ in line-of-sight velocity.  In
some cases, the ``nearest'' neighbor will actually be the
second-nearest, because the SDSS spectrograph cannot simultaneously
take two spectra closer than 55~arcsec; however, the nearest-neighbor
distance $\rpmin$ we calculate is monotonically related to the true
nearest-neighbor distance statistically, at least.  Besides, and as we
show below, for about 45~percent of the SDSS sky region, overlapping
spectroscopic pointings ameliorate or remove this constraint.

As in previous work \citep{quintero04a}, we limit targets to redshifts
$z>0.05$ to mitigate the issues of interpreting spectra taken through
a small ($3$~arcsec diameter) aperture on low-redshift (and thus large
in angle) galaxies.  In addition, when using the clustocentric
distance $\Dcl$, targets are limited to the redshifts $z<0.10$ because
the cluster catalog is limited at redshift $z=0.10$.  When using the
nearest-neighbor distance $\rpmin$, targets are limited to redshifts
$z<0.10$ because redshift $z=0.1$ is the redshift at which a galaxy
with absolute magnitude $M_{^{0.1}i}= -20.0~\mathrm{mag}$ approaches
the flux limit of the SDSS Main Sample.


As mentioned above, there is a technical constraint on the SDSS
spectrographs such that they are unable to simultaneously take spectra
of two galaxies closer than 55~arcsec, which corresponds to proper
transverse distances of 38 and $71\,h^{-1}~\mathrm{kpc}$ at redshifts
0.05 and 0.1 in our standard
$(\Omega_\mathrm{M},\Omega_\mathrm{\Lambda})=(0.3,0.7)$ cosmological
model.  In detail, this constraint (known as ``fiber collisions'')
affects the individual environmental measures of the galaxies in the
sample.  However, the effect is not expected to be large on the
results for several reasons: Only a few percent of SDSS galaxy targets
have been affected by the fiber collisions.  In the mean, our density
estimators are still monotonically related to density at least as well
as other available estimators; the spectrograph constraints mainly
serve to add noise and make the relationship less linear.  The main
results of this work depend on \emph{comparison} of post-starburst and
other galaxies; even if the environmental measures are affected, they
are unable to produce any bias in this comparison (the resolution of
collisions of galaxies in the SDSS Main sample is performed with no
regard to luminosity or color, or any other galaxy property).

Finally, and most importantly, for the results of our environmental
indicator that will be most affected by the spectrosgraph
constraints---the nearest neighbor distance $\rpmin$---we perform the
analysis with the whole sample, and then again with only the
45~percent of the sample that is not affected severely by the
constraint because of multiple coverage by the spectrograph; we find
no significant change.

Of course it is possible to perform environment measurements using the
SDSS imaging data alone \citep[and therefore become free of the
55~arcsec constraint entirely][]{eisenstein03a, hogg03b, blanton05b,
blake04a, goto05a}.  However, these imaging-based environmental
indicators are all either much lower in signal-to-noise \citep[if
background subtraction or deprojection is properly taken into
account][]{eisenstein03a} or else strongly affected by projection
effects (which effectively add noise---probably biased---though it is
difficult to quantify).  For this reason, since we are interested in
very sensitive comparisons of trace populations with dominant
populations, we are much better off with environment indicators using
the full three-dimensional redshift survey data despite its
limitations.

\section{Results}

\begin{figure}
\plotone{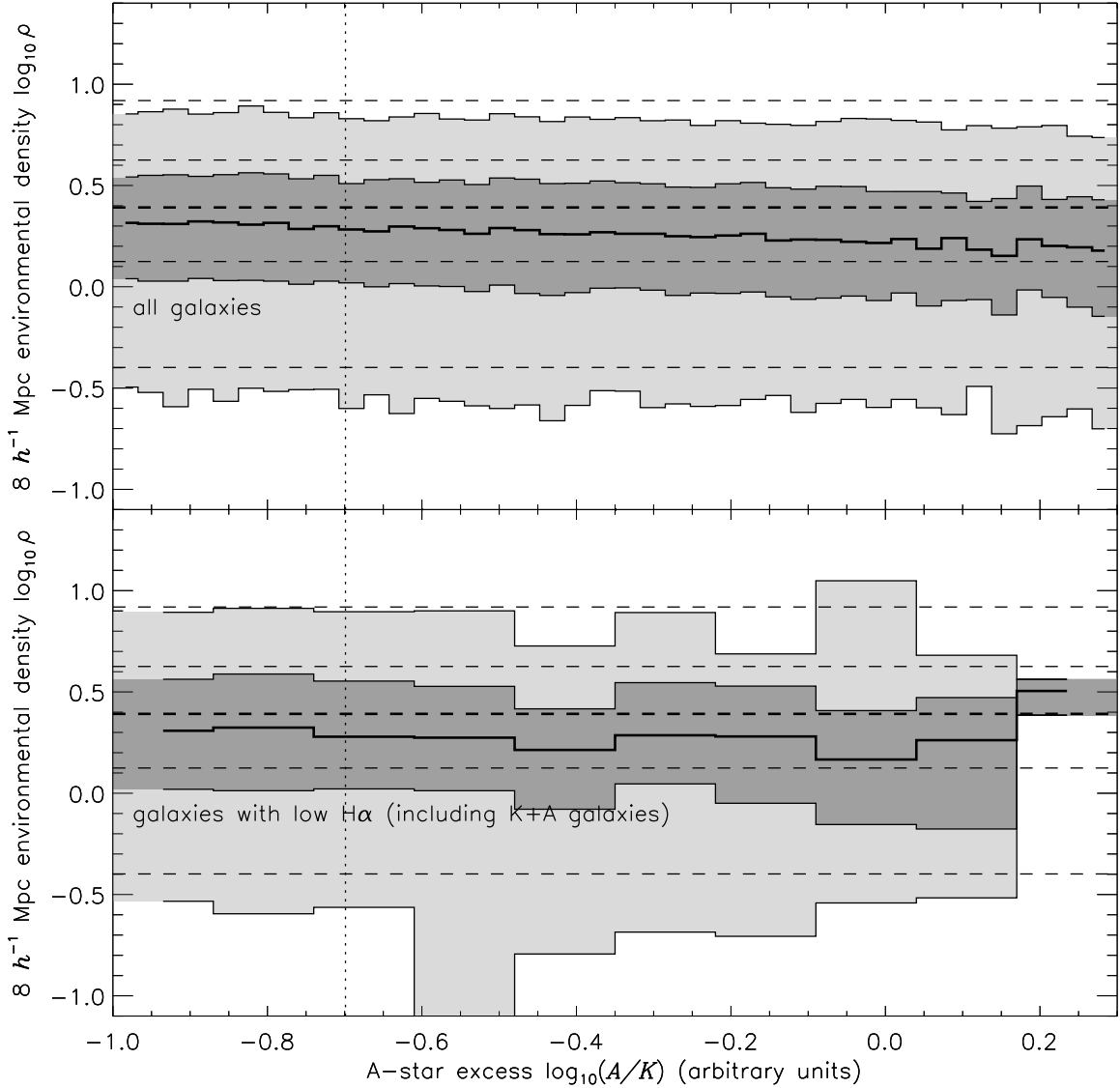}
\caption{The density distribution conditioned on A-star excess.  The
  five lines show the 5, 25, 50, 75, and 95 percentiles in density
  $\rho_8$ (normalized to cosmic mean density) of galaxies in
  $8\,h^{-1}~\mathrm{Mpc}$ radius spheres in comoving redshift space
  around target galaxies as a function of the A-star excess $[A/K]$ in
  the target galaxies.  The top panel is for all galaxies in the
  sample, and the bottom is for those deficient in $\Halpha$ relative
  to the A-star excess (see text for details); the post-starburst or
  K+A galaxies are in the right side of the lower panel.  See text for
  details of the density and A-star excess measurements.  The
  horizontal dashed lines show the same percentiles but for galaxies
  with no A-star excess, and the vertical dotted line is the minimum
  A-star excess required for a galaxy to be classified as star-forming
  (top panel) or K+A (bottom panel) in this work.  Note that both
  populations have very similar dependences of environmental quantiles
  on A-star excess.\label{fig:environments_cond}}
\end{figure}

Figure~\ref{fig:environments_cond} shows the conditional distribution
of environmental density $\rho_8$ as a function of A-star excess
$\log_{10}(A/K)$, in detail, at each A-star excess, it shows the 5th,
25th, 50th, 75th, and 95th percentile in $\rho_8$.  The top panel is
for all galaxies with A-star excesses, and the bottom is for those
characterized as low in $\Halpha$, as described above.  The result is
not just that the median environment is similar for K+A and
star-forming galaxies; the result is that at every value of $A/K$ for
which we have reasonable signal-to-noise, the full distribution of
$\rho_8$ is indistinguishable between the two populations.  This shows
that between $\Halpha$ and $A/K$, it is $A/K$ that best predicts the
large-scale environment in which the galaxy lies.

We have made the equivalents of Figure~\ref{fig:environments_cond} for
the other environment indicators, \ie, transverse clustocentric
distance $\log_{10}(\Dcl/\Rvir)$ and transverse nearest-neighbor
distance $\log_{10}(\rpmin)$.  In both cases, the result is the same
as in Figure~~\ref{fig:environments_cond}: The environment
distribution is the same for star-forming and post-starburst galaxies,
to within the precision of the experiment, at each value of the A-star
excess.

\begin{figure}
\plotone{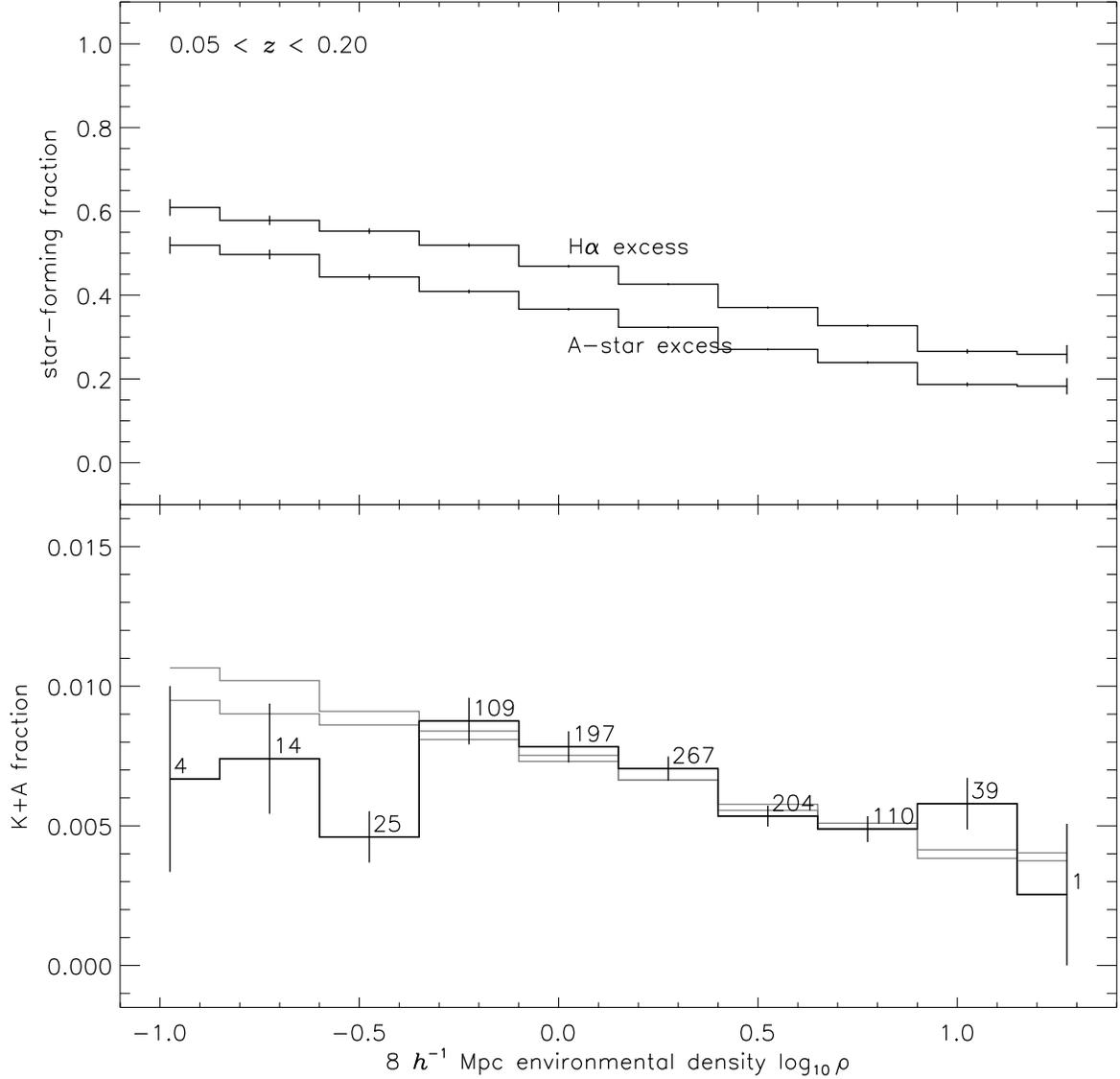}
\caption{The fractional abundance of galaxy populations as a function
  of environmental density.  The top panel shows the fraction of
  galaxies deemed ``star forming'' by $\Halpha$~EW (top curve) and by
  A-star excess $[A/K]$ (bottom curve).  The vertical offset of the
  curves comes from the fact that the A-star excess cut is more severe
  than the $\Halpha$~EW cut (see text for details).  The bottom panel
  shows the same but for the galaxies deemed post-starburst (K+A; see
  text for definition).  Each histogram bin is labeled by the number
  of K+A galaxies in that bin.  Also shown in the bottom panel are
  scaled versions of the curves from the top panel, scaled by the mean
  abundance ratio.\label{fig:environments}}
\end{figure}

The top panel of Figure~\ref{fig:environments} shows the fraction of
galaxies classified as star-forming by the criteria described above,
in the SDSS spectroscopic sample in the redshift range $0.05<z<0.20$,
as a function of environmental density $\rho_8$.  The absolute
fraction is not an interesting number, because it depends on the
luminosity and redshift ranges in the sample, and on the severity of
the star-formation-rate cut (indeed, the fractions show that the
A-star excess cut is more severe).  The bottom panel shows the
fraction of galaxies classified as K+A, with scaled versions of the
curves from the top panel plotted in grey.  It is remarkable how well
the three curves match one another.  The only exception is at very low
densities (recall that the mean density---mean taken over
galaxies---is well above unity because galaxies are clustered), where
there is a slight underdensity of K+A galaxies relative to
star-forming galaxies.

Interestingly, and somewhat beside the point of this work, there is
also a tiny but significant difference visible in
Figure~\ref{fig:environments} in the relative abundances of the two
different kinds of star-forming galaxies ($\Halpha$ excess and A-star
excess) at low environmental density $\rho_8$.  This may have
something to do with the geometry or density or abundance of gas in
the star-forming galaxies in the lowest density environments.

Figure~\ref{fig:clusto} is similar to Figure~\ref{fig:environments}
but with transverse clustocentric distance $\log_{10}(\Dcl/\Rvir)$
acting as the environment indicator.  Again it is remarkable how
similar are the three curves outside of the virial radius.  Inside the
virial radius, the abundance of post-starburst galaxies relative to
star-forming galaxies is higher.  This deviation indicates that a
small number of the post-starburst galaxies may be triggered by
interactions with the intracluster medium.  Unfortunately, the change
in relative abundance is not large, and another equally valid way to
express the discrepancy is that the K+A fraction is a weaker function
of environment than the star-forming fraction.

\begin{figure}
\plotone{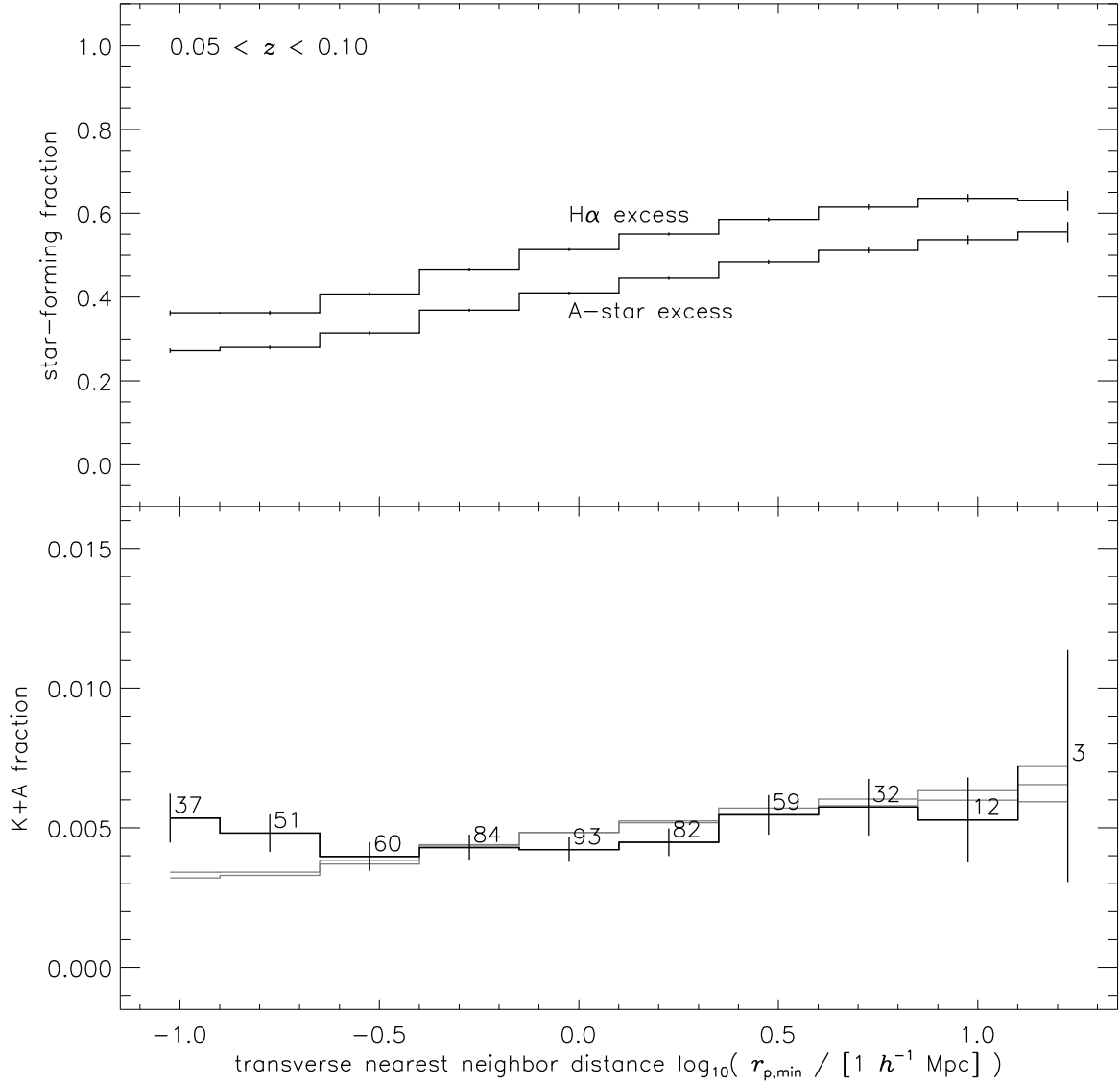}
\caption{Same as Figure~\ref{fig:environments}, except using the
  transverse, proper, nearest-neighbor distance $\rpmin$ (see text) as
  the environment indicator.\label{fig:nearest}}
\end{figure}

Figure~\ref{fig:nearest} is similar to Figure~\ref{fig:environments}
but with transverse nearest-neighbor distance $\rpmin$ acting as the
environment indicator.  Even on these very small scales, the three
curves are very similar.  Again the exception is that there is a very
slight excess of K+As relative to star-formers with nearby neighbors,
or, the K+A fraction is a weaker function of small-scale environment
than the star-forming fraction.  This is consistent with previous
results on very small-scale environments of post-starburst galaxies
\cite{goto05a}.

\begin{figure}
\plotone{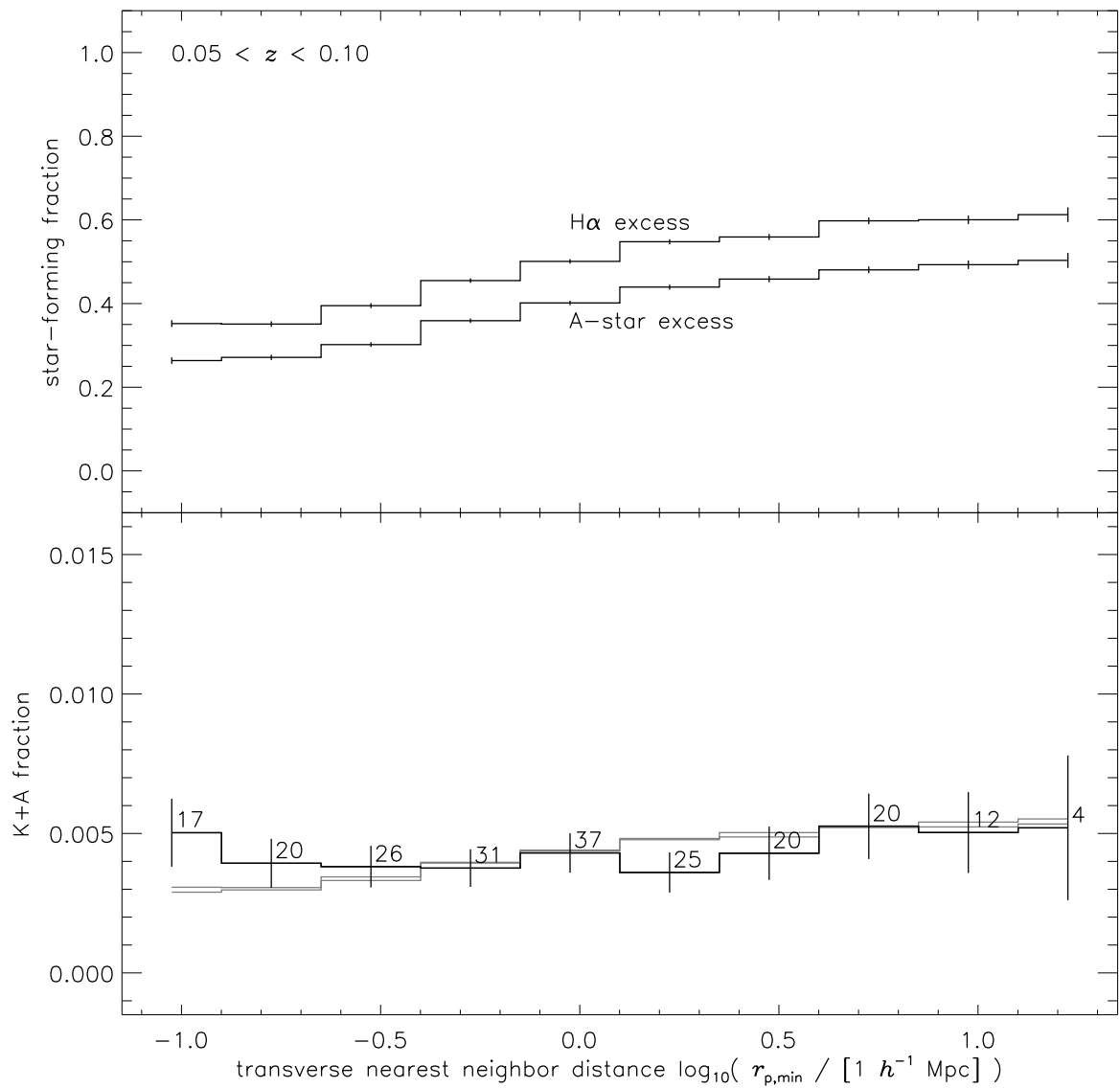}
\caption{Same as Figure~\ref{fig:nearest}, but using only galaxies
  within the SDSS ``overlap'' regions, where the 55~arcsec
  spectrograph constraint does not apply.  The similarity to
  Figure~\ref{fig:nearest} demonstrates that spectrograph constraints
  are not significantly affecting the figure.\label{fig:overlap}}
\end{figure}

As mentioned above, the nearest-neighbor distance $\rpmin$ can in
principle be affected by the SDSS spectrograph 55~arcsec constraint.
For this reason, we have re-made Figure~\ref{fig:environments} using
only galaxies inside SDSS ``overlap regions'' where the spectrograph
had multiple opportunities to obtain spectra in the same sky location,
and therefore ``picked up'' most of the galaxies excluded by the
55~arcsec constraint.  Roughly 45~percent of the sky area used in this
work is inside overlap regions, so the overlap subsample is not small.
Figure~\ref{fig:overlap} is the result; it is the same as
Figure~\ref{fig:nearest} except that it only makes use of galaxies for
which the 55~arcsec constraint is not significant.  The great
similarity of the two Figures demonstrates that our results are not
being strongly affected by this constraint, as expected.

\section{Discussion}

As we have discussed elsewhere \citep{quintero04a}, post-starburst
galaxies plausibly lie on an evolutionary sequence between
disk-dominated galaxies, which are forming stars and contain the
neutral gas fuel for further star formation, and bulge-dominated
galaxies, which have no star-formation fuel and show chemical
signatures of past star-formation bursts.  Post-starburst galaxies
might even be the remnants of major mergers.  A priori, the
environments of these galaxies could be either like those of
disk-dominated galaxies or those of bulge-dominated galaxies, or
somewhere in-between.  Of course, prior to this study, it was already
known that the mean environments of post-starburst galaxies are more
similar to those of disk-dominated galaxies than those of
bulge-dominated galaxies \citep[][Helmboldt et al., in
preparation]{zabludoff96a, quintero04a, blake04a}.  Here we have not
only confirmed this result, we have shown that for each value of the
A-star excess, the post-starburst galaxies with that A-star excess
find themselves in similar environments to star-forming galaxies with
that same A-star excess.  In other words, A-star excess is more
closely related to environment than $\Halpha$~EW, since the
post-starburst galaxies have $\Halpha$~EWs like bulge-dominated
galaxies.

This result confirms our ``null hypothesis'' that the processes that
connect a galaxy's star-formation history to its environment act on
long timescales, longer than A-star lifetimes ($\sim 1~\mathrm{Gyr}$).
This is not surprising, since at typical cosmological velocities
($\sim 100~\mathrm{km\,s^{-1}}$), a galaxy can only travel $\sim
1~\mathrm{Mpc}$ in a Gyr, and there is now pretty good empirical
evidence that everything important about galaxy environments happens
on scales $< 1~\mathrm{Mpc}$ \citep{blanton04c}.

On the other hand, this result is somewhat difficult to reconcile with
the observation from chemical abundances \citep[\eg, ][]{worthey98a,
eisenstein03b} and central stellar densities \citep{quintero04a} that
a typical red (early-type, or bulge-dominated) galaxy has undergone a
brief but strong starburst at some point in its past.  Perhaps the
starbursts are triggered very randomly, with just a slight change in
starburst trigger probability with environment, so we don't see the
relation clearly.  Perhaps the observed relationship between
star-formation rate and environment was set down at very high redshift
and is in fact diluting at the present epoch.  Perhaps the starbursts
are primarily triggered prior to infall into dense environments, in
which case the galaxies somehow ``know'' the environment in which they
will end up!

We have shown that the fraction of the whole SDSS galaxy sample
classified as ``K+A'' (post-starburst) is a function of environment,
and that its dependence on environment is very similar to that of the
fraction of the sample classified as star-forming.  This is also
consistent with our null hypothesis.  The only deviations all have the
sense that the K+A fraction is a slightly weaker function of
environment than the star-forming fraction: We find slightly fewer
K+As in extremely low density environments, and slightly more inside
the virial radii of massive clusters and close to luminous neighbors
than we would expect by naive scaling of the star-forming fraction.

\begin{figure}
\plotone{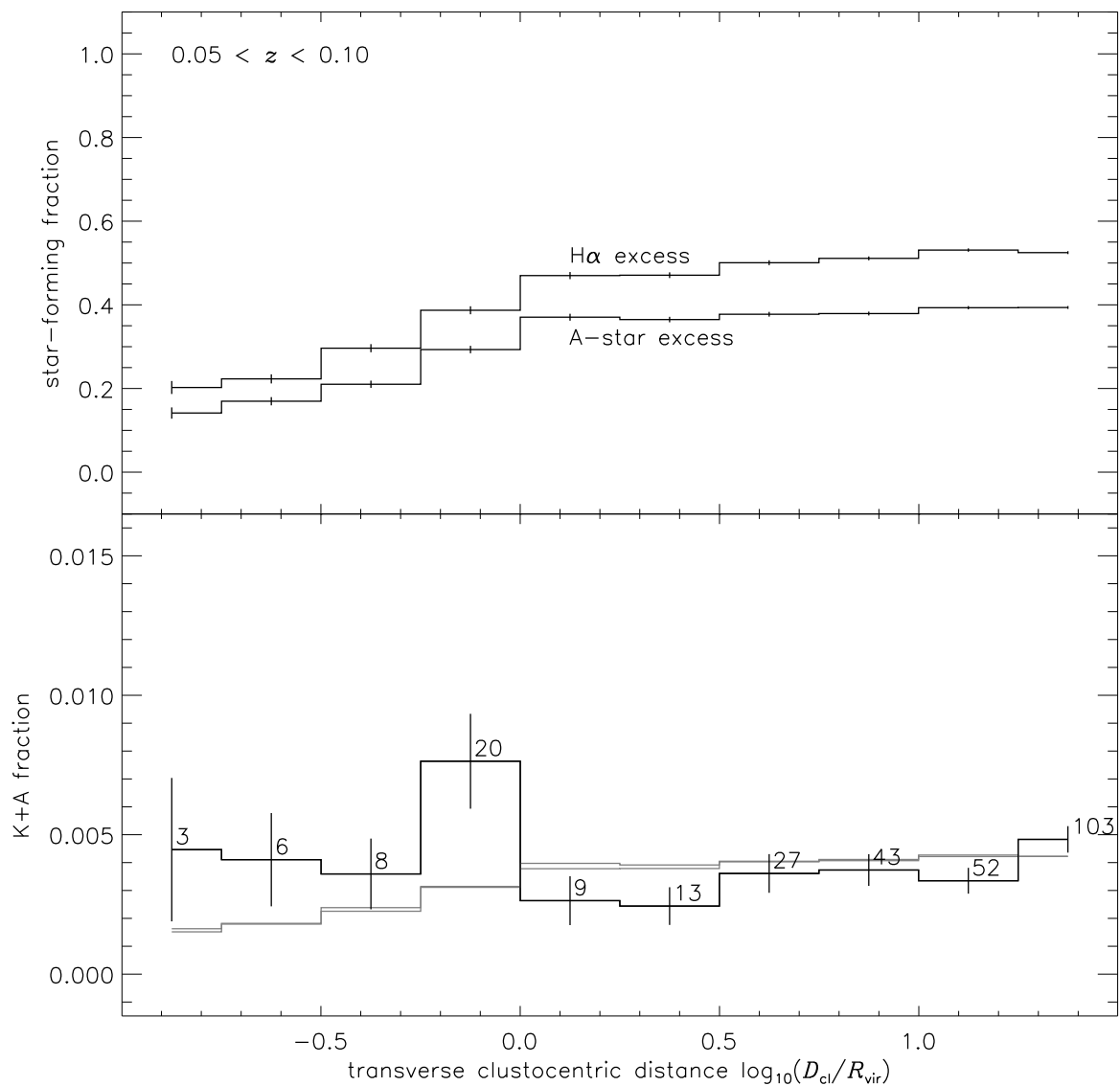}
\caption{Same as Figure~\ref{fig:environments}, except using the
  transverse, virial-scaled clustocentric distance $\Dcl/\Rvir$
  (see text) as the environment indicator.\label{fig:clusto}}
\end{figure}

None of these deviations from the predictions of the null hypothesis
are large, but they may point to triggering mechanisms for the
starburst and post-starburst phases: It is possible that a small
fraction of post-starburst galaxies are triggered by close passages of
luminous galaxies.  It seems likely, from Figure~\ref{fig:clusto},
that a small fraction of post-starbursts are triggered by interactions
with intracluster medium, because the deviation of the post-starburst
fraction relative to the star-forming fraction does occur at the
virial radius.

We can therefore make two negative statements about the triggering of
the \emph{majority} of starbursts, both plausible at the outset: The
first is that only a small fraction of post-starburst galaxies are
triggered by external tidal impulses from close passages of massive
galaxies.  This does not rule out the possibility that they are
created by mergers or accretion events, but if they are, the
truncation of star-formation must occur after the two merging galaxies
are no longer identifiable as separate galaxies.  Our punchline may
seem to be at odds with the punchline of some previous work
\citep{goto05a}.  Quantitatively, however, there is no disagreement,
given the differences in methodology, and in fact both studies do find
small excesses of neighbors at small scales.  This suggests that
although tidal-impulse triggering (\ie, triggering by close passages
of massive neighbors) no-doubt occurs, it is not the dominant
mechanism.

The second negative statement we can make is that only a small
fraction of post-starburst galaxies are created by IGM interactions on
infall into clusters.  Certainly some are; the identification of the
virial radius in the abundance of post-starburst galaxies relative to
star-forming galaxies is intriguing.  The galaxy populations inside
clusters are very different in morphological mix and
star-formation-history mix than galaxy populations elsewhere.  What
physical process are involved in enforcing these differences?  Many
have hypothesized---quite naturally---that radical transformations
must happen on infall \citep{poggianti99a, balogh00a, kodama01a};
indeed some galaxies have been ``caught in the act'' of a radical
transformation \citep[\eg, ][]{vogt04a}.  Quantitatively, however, the
fraction of galaxies showing clear evidence for intracluster medium
interactions is very small.  This is consistent with prior work in
this area \citep{poggianti99a, balogh00a, kodama01a, vogt04a}.

The orbital time for a galaxy falling into a cluster is only a
few times longer than the lifetime of A stars, so the excess of
post-starburst galaxies in the infall regions is expected to ``smear''
into the cluster center; \ie, the morphology of
Figure~\ref{fig:clusto} is about right (given the small numbers)
for the intra-cluster medium hypothesis.  What remains to be seen is
whether, quantitatively, the tiny number of post-starburst galaxies
observed within clusters is consistent with the competing facts that
clusters are continuously growing by accretion of field galaxies and
groups and that the morphological and star-formation mix in clusters
is different inside and outside of clusters; the prodigious expected
infall means that a lot of galaxies ought to be changing their
morphologies and star-formation rates on infall.  We expect such an
analysis---beyond the scope of this paper---to conclude that either
the transformation process is slow ($>1$~Gyr, which is not necessarily
consistent with observations of abundances in bulge-dominated
galaxies), or else that the galaxies in clusters somehow ``knew in
advance'' (from their pre-infall group or filament environment) that
they were destined to end up in a cluster.  A philosophical point
arises here: The fact that galaxy populations inside clusters are
different from those outside does not \emph{necessarily} mean that
transformations take place as galaxies move from one environment to
the other.

The remaining hypotheses for the triggering of the starburst (or, more
properly, star-formation truncation) events that precedes the
post-starburst phases of these galaxies are: some kinds of random
internal catastrophes or some kinds of galaxy--galaxy mergers.  This
latter possibility, which is consistent with all of the results here,
is directly supported by the discovery of post-merger morphological
signatures (\eg, tidal arms) in many post-starburst galaxies
\citep{yang04a, goto05a}.  It is also exciting, because merging is one
of the fundamental processes of cosmogony, and holds great promise for
providing precise connections between cosmological observations and
theory at small scales.

\acknowledgments It is a pleasure to thank Tomo Goto, Joe Helmboldt,
Don Schneider, Michael Strauss, Douglas Tucker, and an anonymous
referee for useful comments on the manuscript.  DWH and MM were
generously hosted by the Massachusetts Institute of Technology Kavli
Institute for Astrophysics and Space Research during the period of
this research.  DWH, MM, MRB, and ADQ are partially supported by the
National Aeronautics and Space Administration (NASA; grant
NAG5-11669); DWH, MM, and MRB are partially supported by the National
Science Foundation (NSF; grant AST-0428465).  This research made use
of the NASA Astrophysics Data System.  It also made use of the
``idlutils'' codebase maintained by David Schlegel, Doug Finkbeiner,
and others.

Funding for the creation and distribution of the SDSS Archive has been
provided by the Alfred P. Sloan Foundation, the Participating
Institutions, NASA, the NSF, the U.S. Department of Energy, the
Japanese Monbukagakusho, and the Max Planck Society.  The SDSS Web
site is ``http://www.sdss.org/''.

The SDSS is managed by the Astrophysical Research Consortium for
the Participating Institutions.  The Participating Institutions are
The University of Chicago, Fermilab, the Institute for Advanced Study,
the Japan Participation Group, The Johns Hopkins University, the
Korean Scientist Group, Los Alamos National Laboratory, the
Max-Planck-Institute for Astronomy, the Max-Planck-Institute
for Astrophysics, New Mexico State University, University of
Pittsburgh, University of Portsmouth, Princeton University, the United
States Naval Observatory, and the University of Washington.

\clearpage

\clearpage

\clearpage

\clearpage

\clearpage

\clearpage

\end{document}